\documentclass[seceq]{ptptex}

\usepackage{graphicx}

\newcommand{\bc}{\begin{center}}
\newcommand{\ec}{\end{center}}
\newcommand{\be}{\begin{equation}}
\newcommand{\ee}{\end{equation}}
\newcommand{\bea}{\begin{eqnarray}}
\newcommand{\eea}{\end{eqnarray}}

\title{
What atomic liquids can teach us about quark liquids%
}

\author{
Thomas \textsc{Sch\"afer}\footnote{ e-mail address:
thomas\_schaefer@ncsu.edu}  }

\inst{
Department of Physics, North Carolina State University,
Raleigh, NC 27695
}

\abst{
We discuss some aspects of cold atomic Fermi gases in the unitarity 
limit that are of interest in connection with the physics of quark 
matter and the quark gluon plasma. We consider, in particular, 
the equation of state, transport properties, the critical temperature 
for pair condensation, and the response to a pair breaking stress.}

\begin{document}

\maketitle

\section{Introduction}
\label{sec_intro}

  Over the last ten years there has been truly remarkable progress 
in the study of cold, dilute gases of fermionic atoms in which the 
scattering length $a$ of the atoms can be controlled experimentally. 
These systems can be realized in the laboratory using Feshbach 
resonances, see\cite{Regal:2005} for a review. A small negative 
scattering length corresponds to a weak attractive interaction between 
the atoms. This case is known as the BCS (Bardeen-Cooper-Schrieffer)
limit. As the strength of the interaction increases the scattering 
length becomes larger. It diverges at the point where a bound state 
is formed. The point $a=\infty$ is called the unitarity limit, because 
the scattering cross section saturates the $s$-wave unitarity bound 
$\sigma=4\pi/k^2$. On the other side of the resonance the scattering 
length is positive. In the BEC (Bose-Einstein condensation) limit the 
interaction is strongly attractive and the fermions form deeply bound 
molecules.

 The unitarity limit is of particular interest. In this limit the 
atoms form a strongly coupled quantum liquid which exhibits universal
behavior. In the BCS limit the atomic gas is characterized by the 
small parameter $(k_Fa)$, where $k_F$ is the Fermi momentum. In the
unitarity limit this parameter is infinite and the system is strongly 
coupled. Universality arises from the fact that short distance 
effects are suppressed by the small parameter $(k_Fr)$, where $r$ is 
the effective range. In the experiments performed to date $(k_Fr)\ll 1$,
and we expect that the theoretical limit $(k_Fa)\to\infty$, $(k_Fr)\to 0$ 
is well defined. Dilute fermions in the unitarity limit provide 
an interesting model system in which a number of questions regarding 
the behavior of strongly coupled quantum liquids can be studied. In 
this contribution we shall study a number of issues that are of 
relevance to the phase diagram of QCD:

\begin{itemize}

\item What is the equation of state at strong coupling? Is the
transition from weak to strong coupling smooth?

\item What are the transport properties at strong coupling? Are 
transport properties more sensitive to the coupling than thermodynamic
quantities? Do atomic liquids respect the proposed bound on 
the ratio of shear viscosity to entropy density? 

\item What is the critical temperature for pairing? Is there 
a universal upper bound on $T_c/E_F$, where $T_c$ is the 
critical temperature and $E_F$ is the Fermi energy.

\item How does the paired state below $T_c$ respond to a 
pair breaking stress? Are there any intermediate states
that separate the fully paired state from the normal state?

\end{itemize}

\section{Equation of State}
\label{sec_eos}

 Asymptotic freedom implies that the equation of state of a 
quark gluon plasma at $T\gg\Lambda_{QCD}$ is that of a free
gas of quarks and gluons. Numerical results from lattice QCD 
calculations show that at $T\sim 2T_c$, which is relevant to 
the early stages of heavy ion collisions at RHIC, the pressure
and energy density reach about 85\% of the free gas limit. This 
is consistent with the first order perturbative correction. 
Higher order terms in the perturbative expansion are very poorly 
convergent, but this problem can be addressed using resummation 
techniques\cite{Blaizot:2003tw}. In this framework the degrees 
of freedom are dressed quasi-quarks and quasi-gluons, and these 
quasi-particles are weakly interacting.

 Transport properties of the plasma indicate that this may not 
be correct. Experiments at RHIC indicate that the viscosity of the
plasma is very small, and that the opacity for high energy jets
is very large. An interesting perspective on this 
issue is provided by a strong coupling calculation performed 
in the large $N_c$ limit of ${\cal N}=4$ SUSY Yang Mills theory. 
The calculation is based on the duality between the strongly coupled 
gauge theory and weakly coupled string theory on $AdS_5\times S_5$ 
discovered by Maldacena\cite{Maldacena:1997re}. The correspondence 
can be extended to finite temperature. In this case the relevant 
configurations is an $AdS_5$ black hole. The temperature of the 
gauge theory is given by the Hawking temperature of the black hole, 
and the entropy is given by the Hawking-Beckenstein formula. The result
is that the entropy density of the strongly coupled field theory
is equal to 3/4 of the free field theory value\cite{Gubser:1998nz}.
This implies that thermodynamics is not drastically effected
in going from weak to strong coupling.

 The crossover from weak to strong coupling can also be studied 
in the context of cold atomic gases. In the non-interacting limit 
the energy per particle is given by $E/N=3E_F/5$. The Fermi energy 
$E_F=k_F^2/(2m)$ is related to the density $N/V=k_F^3/(3\pi^2)$. 
In the BCS limit interactions reduce the energy per particle. To 
leading order in $(k_Fa)$ we have 
\be 
\frac{E}{N}= \frac{3E_F}{5}\left\{ 1 + \frac{10}{9\pi}
  (k_Fa) + \ldots \right\} .
\ee
In the unitarity limit $(k_Fa)\to\infty$ the energy per particle 
must be a universal constant times the free Fermi gas result, 
$E/N = \xi (3E_F/5)$. The calculation of the dimensionless 
quantity $\xi$ is a non-perturbative problem. The most accurate
results for $\xi$ are believed to come from Green Function Monte 
Carlo (GFMC) calculations. Carlson et al.~find\cite{Carlson:2003}
$\xi=0.44$. This value is consistent with recent experimental 
determinations. GFMC calculations also show that the crossover 
from weak to strong coupling is smooth. 

 It is interesting to find analytical approaches to the equation 
of state in the unitarity limit. We have recently summarized this 
subject in\cite{Schafer:2006ky}. There are a number of methods that 
can be systematically improved:

\begin{itemize}
\item An expansion in the number of species\cite{Sachdev:2006}.
 At leading order this is essentially the BCS approximation, 
 and higher orders take into account fluctuations around the 
 BCS mean field. The result is $\xi=0.59+O(1/N)$.

\item An epsilon expansion\cite{Nussinov:2004,Nishida:2006br} around 
$d=4$ or $d=2$. The $4-\epsilon$ expansion involves weakly coupled 
bosons and fermions, while the $2+\bar\epsilon$ expansion is related
to the perturbative $(k_Fa)$ expansion. The most reliable results are 
obtained by combining the two methods. Arnold\cite{Arnold:2006fr} 
et al.~conclude that $\xi=(0.30-0.37)$. 

\item An expansion in one over the number of spatial 
dimensions\cite{Steele:2000qt,Schafer:2005kg}. This method
corresponds to the hole line expansion of Bethe and Brueckner. 
The pair condensation energy is formally suppressed by $1/d$. 
The leading order result is $\xi=1/2+O(1/d)$.
\end{itemize}

 The various methods emphasize different aspects of the physics 
of a cold fermion gas, and they all have their advantages and 
disadvantages. None of them appear to converge rapidly. 

\section{Transport Properties}
\label{sec_trans}

 The matter produced at RHIC is characterized by strong radial 
and elliptic flow\cite{rhic:2005}. This observation has lead to
the conclusion that the shear viscosity to entropy density ratio
of the quark gluon plasma at temperatures near $T_c$ must be 
very small\cite{Teaney:2003kp}, $\eta/s\ll 1$. In the weak coupling 
limit the shear viscosity can be computed in perturbative QCD. The 
result is\cite{Arnold:2003zc}
\be 
\frac{\eta}{s} = \frac{5.12}{g^4\log(2.42 g^{-1})}.
\ee
In a weakly coupled ($g\sim 1$) QGP $\eta/s$ is very large. The 
shear viscosity to entropy density ratio decreases as the coupling 
increases, but it is hard to extrapolate the weak coupling result 
to the strong coupling domain. Kovtun et al.~conjectured that there 
is a universal lower bound\cite{Kovtun:2004de} $\eta/s\geq \hbar/
(4\pi k_B)$. The bound is saturated in the case of strongly coupled 
gauge theories that have a gravity dual, like the ${\cal N}=4$ 
SUSY gauge theory discussed in Sect.~\ref{sec_eos}.

 In cold atomic gases we can reliably compute $\eta/s$ in the BCS 
limit. The ratio is temperature dependent and has a minimum at $T
\sim T_F$, where $T_F=E_F/k_B$ is the Fermi temperature. The shear 
viscosity is proportional to $1/a^2$, and $\eta/s$ is very large in 
the weak coupling limit. As in the QCD case, there are no controlled 
strong coupling calculations. It is possible, however, to reliably 
extract $\eta/s$ from experimental data on the damping of collective 
oscillations\cite{Kavoulakis:1998}. Collective modes have been studied 
in a number of experiments\cite{Kinast:2004}. In the unitarity limit 
the frequency of collective modes is well described by ideal 
hydrodynamics\cite{Heiselberg:2004}. The energy dissipated due to 
viscous effects is 
\be
\dot{E} = -\frac{1}{2} \int d^3x\, \eta
  \left(\partial_iv_j+\partial_jv_i-\frac{2}{3}\delta_{ij}
      \partial_k v_k \right)^2  
   - \int d^3x \; \zeta \big( \partial_iv_i\big)^2 \, , 
\ee
where $v_i$ is the flow velocity, $\eta$ is the shear viscosity and 
$\zeta$ is the bulk viscosity. In the unitarity limit the system is 
scale invariant and the bulk viscosity in the normal phase vanishes. 

\begin{figure}[t]
\begin{center}
\includegraphics[width=8cm]{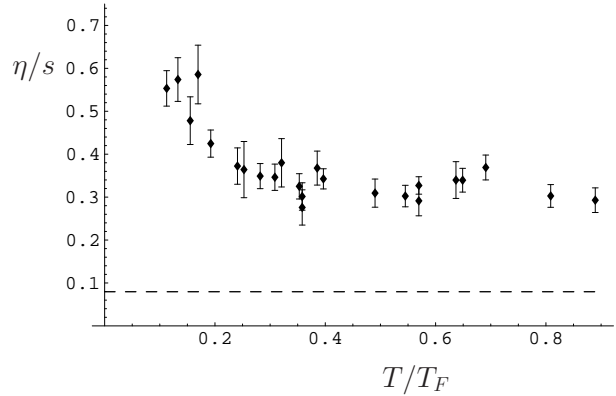}
\end{center}
\caption{\label{fig_eta_s}
Viscosity to entropy density ratio of a cold atomic gas in the 
unitarity limit. This plot is based on the damping data 
published in \cite{Kinast:2005} and the thermodynamic data in 
\cite{Kinast:2005b,Luo:2006}. The dashed line shows the 
conjectured viscosity bound $\eta/s=1/(4\pi)$. }
\end{figure}

 We recently analyzed the experimental data of the Duke 
group\cite{Schafer:2007pr}. Kinast et al.~measure the damping rate 
of the radial breathing mode in an elongated, axially symmetric
trap \cite{Kinast:2005}. They report measurements of the damping
rate $\Gamma$ in units of the radial trap frequency as a function
of $T/T_F$. The shear viscosity to entropy density ratio is given
by 
\be 
\label{eta_s}
\frac{\eta}{s}  =\frac{3}{4}\, \xi^{1/2} (3N)^{1/3} 
 \left(\frac{\Gamma}{\omega_\perp}\right)
 \left(\frac{\bar\omega}{\omega_\perp}\right)
 \left(\frac{N}{S}\right),
\ee
where $\bar\omega=(\omega_\perp^2\omega_z)^{1/3}$ is the geometric
mean of the trap frequencies, $N$ is the number of atoms, and 
$S/N$ is the entropy per particle. Our results are shown in 
Fig.~\ref{fig_eta_s}. We observe that $\eta/s$ has a shallow
minimum near $T_c\sim T_F/3$. The value at the minimum is 
$\eta/s\sim 1/3$, roughly four times bigger than the proposed
bound.

\section{Critical Temperature}
\label{sec_pair}

 In the limit of high baryon density and low temperature the QCD 
phase diagram contains a number of color superconducting phases. 
Color superconductivity is characterized by the formation
of quark Cooper pairs. At asymptotically large density the 
attraction is due to one-gluon exchange. In this limit the 
pairing gap is given by\cite{Son:1999uk}
\be
\label{gap_oge}
\Delta = 2\Lambda_{BCS}
   \exp\left(-\frac{\pi^2+4}{8}\right)
   \exp\left(-\frac{3\pi^2}{\sqrt{2}g}\right).
\ee
where $g$ is the running coupling constant evaluated at the scale 
$\mu$ and $\Lambda_{BCS}=256\pi^4(2/N_f)^{5/2}g^{-5}\mu$. Here,
$\mu$ is the baryon chemical potential and $N_F$ is the number 
of flavors. This result exhibits a non-BCS like dependence on the 
coupling constant which is related to the presence of unscreened
magnetic gluon exchanges. The critical temperature is nevertheless
given by the BCS result $T_c=e^\gamma\Delta/\pi$. 

 In the weak coupling limit the gap and the critical temperature are 
exponentially small. The ratio $T_c/E_F$ increases with $g$ and reaches 
a maximum of $T_c=0.025E_F$ at $g=4.2$. The maximum occurs at strong 
coupling and the result is not reliable. Using phenomenological 
interactions, or extrapolating the QCD Dyson-Schwinger equations 
into the strong coupling domain \cite{Nickel:2006vf}, one finds 
critical temperatures as large as $T_c=0.15E_F$. 

 At low temperature the atomic gas becomes superfluid. If the 
coupling is weak then the gap and the critical temperature can be 
calculated using BCS theory. The result is 
\be
\label{gap_bcs}
\Delta = \frac{8E_F}{(4e)^{1/3}e^2}
   \exp\left(-\frac{\pi}{2k_F|a|}\right),
\ee
where the factor $(4e)^{1/3}$ is the screening correction first 
computed by Gorkov et al.~\cite{Gorkov:1961}. Higher order 
corrections are suppressed by powers of $(k_Fa)$. In BCS theory 
the critical temperature is given by $T_c=e^\gamma\Delta/\pi$. 
Clearly, the critical temperature grows with the scattering length. 
Naively extrapolating equ.~(\ref{gap_bcs}) to the unitarity limit 
gives $T_c\simeq 0.28 E_F$. The value of $T_c$ has been determined 
in a number of Monte Carlo calculations. Burovski et al.~
find\cite{Burovski:2006} $T_c=0.152(7)E_F$, while Bulgac et 
al.~obtain\cite{Bulgac:2005pj} $T_c=0.23(2)E_F$ and Akkinei et 
al.~quote\cite{Akkinei:2006} $T_c=0.25 E_F$. The larger values 
of $T_c$ are in better agreement with the transition observed 
in trapped systems\cite{Bulgac:2007ah}.

\section{Stressed Pairing}
\label{sec_stress}

 The exact nature of the color superconducting phase in QCD depends
on the baryon chemical potential, the number of quark flavors and 
on their masses. If the baryon chemical is much larger than the 
quark masses then the ground state of QCD with three flavors is the 
color-flavor-locked (CFL) phase. The CFL phase is characterized by 
the pair condensate \cite{Alford:1999mk}
\be
\label{cfl}
 \langle \psi^a_i C\gamma_5 \psi^b_j\rangle  =
  (\delta^a_i\delta^b_j-\delta^a_j\delta^b_i) \phi .
\ee
This condensate leads to a gap in the excitation spectrum
of all fermions and completely screens the gluonic interaction.
Both the chiral $SU(3)_L\times SU(3)_R$ and color $SU(3)$
symmetry are broken, but a vector-like $SU(3)$ flavor symmetry
remains unbroken. In the real world the strange quark mass is 
not equal to the masses of the up and down quark and flavor 
symmetry is broken. At high baryon density the effect of the 
strange quark mass is governed by the shift $\mu_s = m_s^2/(2\mu)$ 
of the strange quark Fermi energy.

 The response of the CFL state to a non-zero $\mu_s$ is a 
difficult problem that has not been fully resolved, even in 
the weak coupling limit. There are three energy scales that 
are important 

\begin{itemize}
\item $m_K\sim (m_um_s)^{1/2}(\Delta/\mu)\ll \Delta$ is the 
mass of the neutral strange Goldstone boson, the $K^0$. When
$\mu_s>m_K$ the CFL phase undergoes a transition to a phase
with kaon condensation\cite{Bedaque:2001je}.

\item $\mu_s^{(1)}\sim \Delta$ is the critical value of $\mu_s$
at which the first fermion mode becomes gapless. For $\mu_s>
\mu_s^{(1)}$ the CFL phase (with or without kaon condensation)
is a gapless superfluid\cite{Alford:2003fq}.

\item $\mu_s^{(2)}\sim 2\Delta$ is the critical value of $\mu_s$
beyond which CFL pairing breaks down completely. For $\mu_s>
\mu_s^{(2)}$ the CFL phase is replaced by a less symmetric phase,
like the 2SC phase or single flavor pairing in the spin-one
channel.

\end{itemize}

 The most difficult part of the phase diagram is the region
$\mu_s^{(1)}<\mu_s<\mu_s^{(2)}$. Gapless fermion modes cause
instabilities in the superfluid density and the magnetic screening 
masses\cite{Huang:2004bg}. Near $\mu_s^{(1)}$ this instability can 
be resolved by a small Goldstone boson current\cite{Kryjevski:2005qq}. 
Closer to $\mu_s^{(2)}$ the Goldstone boson current may become large, 
and multiple currents can appear. In this limit the ground state is 
more appropriately described as a LOFF phase\cite{Casalbuoni:2005zp}. 
We shall describe the LOFF state in more detail below. 

\begin{figure}[t]
\bc\includegraphics[width=7.5cm]{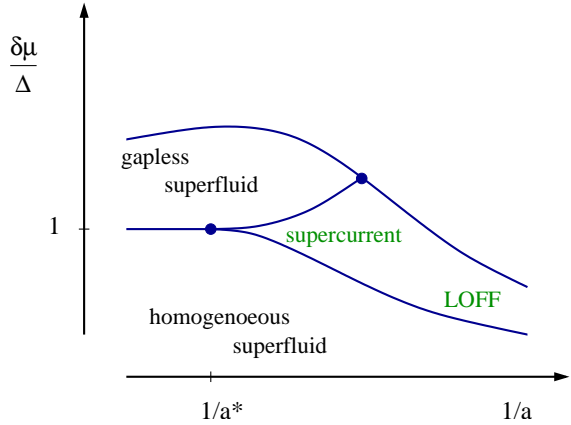}\ec
\caption{\label{fig_ph}
Conjectured phase diagram for a polarized cold atomic Fermi gas
as a function of the scattering length $a$ and the difference 
in the chemical potentials $\delta\mu=\mu_\uparrow-\mu_\downarrow$, 
from Son \& Stephanov (2005).}
\end{figure}

 The atomic superfluid involves equal numbers of spin up and spin 
down fermions. The physical situation which is analogous to the
response of the CFL phase to $m_s$ is the response of the 
atomic superfluid to a non-zero chemical potential coupled to 
the third component of spin, $\delta\mu=\mu_\uparrow-\mu_\downarrow$.
A conjectured (and, most likely, oversimplified) phase diagram 
for a polarized gas is shown in Fig.~\ref{fig_ph}. In the BEC 
limit the gas consists of tightly bound spin singlet molecules. 
Adding an extra up or down spin requires energy $\Delta$. For 
$|\delta\mu|>\Delta$ the system is a homogeneous mixture of a Bose 
condensate and a fully polarized Fermi gas. This mixture is stable 
with respect to phase separation.

 In the BCS limit the problem was first analyzed by Larkin, Ovchninikov, 
Fulde and Ferell (LOFF) \cite{Larkin:1964}. Consider the homogeneous 
solutions to the BCS gap equation for $\delta\mu\neq 0$. In the regime 
$\delta\mu<\Delta_0$ where $\Delta_0=\Delta(\delta\mu \!=\!0)$ the gap 
equation has a solution with gap parameter $\Delta=\Delta_0$. This 
solution is stable if $\delta\mu<\Delta_0/\sqrt{2}$ but only meta-stable 
in the regime $\Delta_0/\sqrt{2}<\delta\mu<\Delta_0$. The BCS solution 
has vanishing polarization. The transition to a polarized normal phase 
is first order, and systems at intermediate polarization correspond 
to mixed phases.

LOFF~studied whether it is possible to find a stable solution in 
which the gap has a spatially varying phase
\be 
\label{loff}
 \Delta(\vec{x})= \Delta e^{2i\vec{q}\cdot\vec{x}}.
\ee
This solution exists in the LOFF window $\delta\mu_1<\delta\mu<
\delta\mu_2$ with $\delta\mu_1=\Delta_0/\sqrt{2}\simeq 0.71
\Delta_0$ and $\delta\mu_2\simeq 0.754\Delta_0$. The LOFF momentum 
$q$ depends on $\delta\mu$. Near $\delta\mu_2$ we have $qv_F
\simeq 1.2\delta\mu$, where $v_F$ is the Fermi velocity. The gap 
$\Delta$ goes to zero near $\delta\mu_2$ and reaches $\Delta\simeq 
0.25\Delta_0$ at $\delta\mu_1$.

  These results suggest that for some value of the scattering length
between the BEC and BCS limits the homogeneous superfluid becomes 
unstable with respect to the formation of a non-zero supercurrent 
$\vec\nabla\varphi$, where $\varphi$ is the phase of the condensate. 
We can study the onset of the instability using the effective lagrangian
\be
\label{leff_gbcs}
  {\cal L} =  \psi^\dagger \Big(i\partial_0 - \epsilon(-i\vec{\partial}) 
  - i(\vec{\partial}\varphi)\cdot
   \frac{\stackrel{\scriptstyle\leftrightarrow}{\partial}}{2m} 
    \Big)\psi
  + \frac{f_t^2}{2} \dot\varphi^2 - \frac{f^2}{2} (\vec{\partial}\varphi)^2.
\ee
Here, $\psi$ describes a fermion with dispersion law $\epsilon(\vec{p})$ 
and $\varphi$ is the superfluid Goldstone mode. The low energy parameters 
$f_t$ and $f$ are related to the density and the velocity of sound. Setting 
up a current $\vec{v}_s=\vec{\partial}\varphi/m$ requires energy $f^2m^2
v_s^2/2$. The fermion dispersion law in the presence of a non-zero current 
is $\epsilon_v(\vec{p})=\epsilon(\vec{p})+\vec{v}_s\cdot\vec{p}-\delta\mu$, 
and a current can lower the energy of the fermions. The free energy 
functional was analyzed by Son and Stephanov \cite{Son:2005qx}. They
noticed that the stability of the homogeneous phase depends crucially on 
the nature of the dispersion law $\epsilon(p)$. In the BEC limit the 
minimum of the dispersion curve is at $p=0$ and there is no current 
instability. In the BCS limit the minimum is at $p\neq 0$, the fermion 
contribution is amplified by a finite density of states on the Fermi 
surface, and the system is unstable with respect to the formation of a 
non-zero current. 

 There is an ongoing effort dedicated to the experimental study 
of the phase diagram as a function of scattering length, polarization, 
and temperature. We cannot adequately summarize all of these experiments 
here. The MIT group has mapped out the superfluid-normal transition 
line in the $(\delta\mu,a)$ plane\cite{Zwierlein:2006}. In the 
unitarity limit the transition occurs at a population imbalance 
$(n_\uparrow-n_\downarrow)/(n_\uparrow+n_\downarrow)\simeq 70$ \%. 
Currently, there is no evidence for inhomogeneous or anisotropic
states like the LOFF phase. 

Acknowledgments: This work is supported in part by the 
US Department of Energy grant DE-FG02-03ER41260.


\end{document}